\documentclass[a4paper]{article}

\usepackage[utf8]{inputenc} 
\usepackage[T1]{fontenc}    
\usepackage{hyperref}       
\usepackage{url}            
\usepackage{booktabs}       
\usepackage{amsfonts}       
\usepackage{nicefrac}       
\usepackage{graphicx}       
\graphicspath{{media/}}     

\usepackage{graphicx}
\usepackage{titlesec}
\usepackage{geometry}
\usepackage{amsmath}
\usepackage{csquotes}
\usepackage{amssymb}
\newgeometry{vmargin={25mm,50mm}, hmargin={35mm,35mm}}
\titleformat{\paragraph}{\normalfont\normalsize\bfseries}{\theparagraph}{1em}{}
\titlespacing*{\paragraph}{0pt}{3.25ex plus 1ex minus .2ex}{1.5ex plus .2ex}

\usepackage{fancyhdr}

\fancypagestyle{plain}{
\fancyhf{}
\pagestyle{fancy}
\fancyhead[C]{\includegraphics[height=2cm]{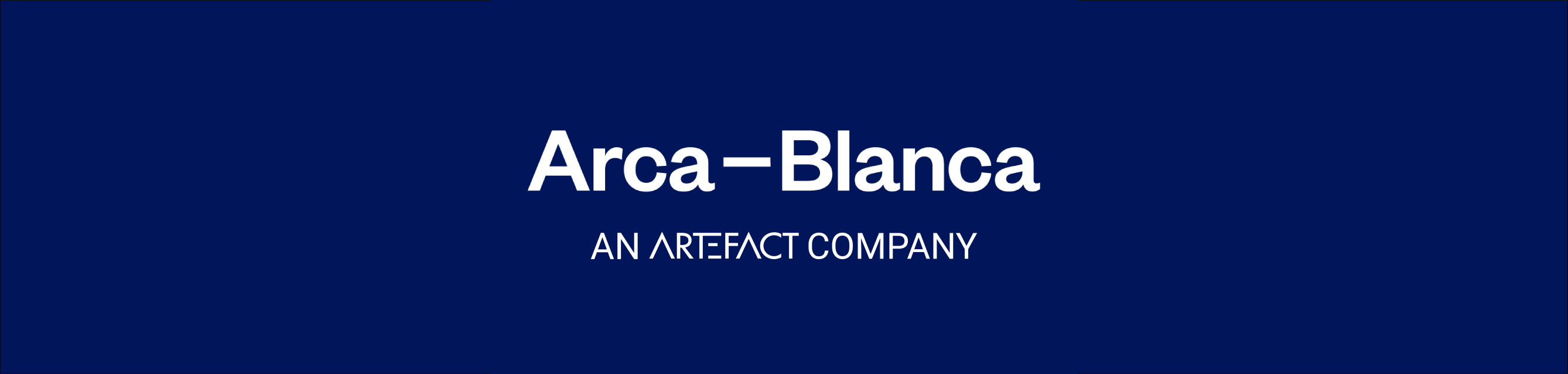}}
\fancyfoot[L]{Copyright \textcopyright \vspace{6pt} 2023 Arca Blanca}
}
\usepackage{amsthm}

\fancypagestyle{AB}{
\fancyhf{}
\pagestyle{fancy}
\fancyhead[L]{\includegraphics[height=1cm]{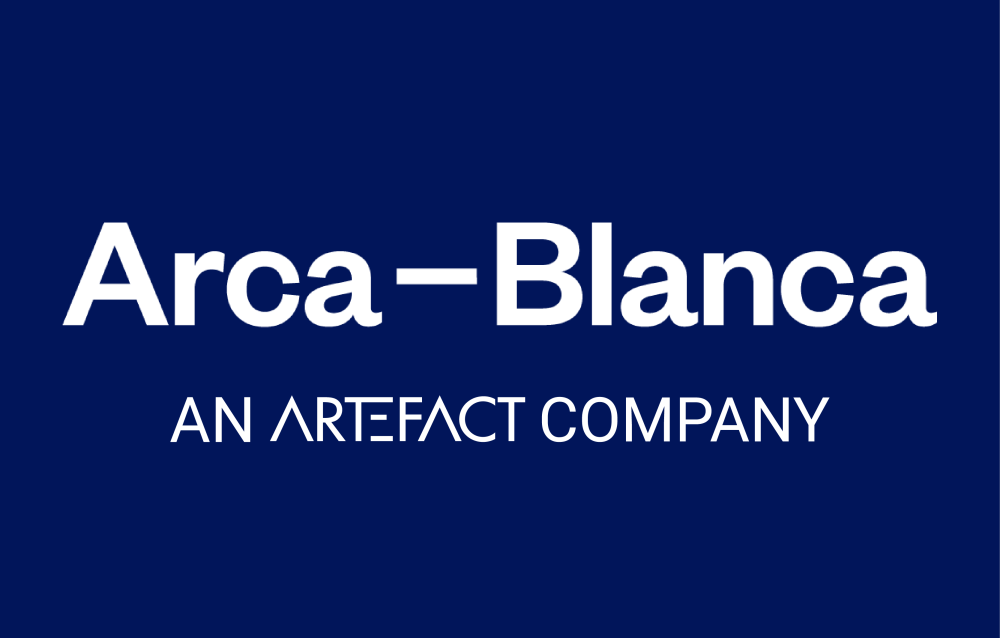}}
\fancyfoot[L]{Copyright \textcopyright \vspace{6pt} 2023 Arca Blanca}
\fancyfoot[R]{ \thepage}
}
\usepackage{authblk}
\title{Instability of Deep Learning models is not a problem, but a result of our framing of classification problems}
\author{  
  Rozario, Savio\\
  \texttt{savio.rozario@arcablanca.com}
  \and
  \v{C}evora, George\\
  \texttt{george.cevora@arcablanca.com}
  }
\affil{Arca Blanca Ltd.\\Manfield House, One Southampton St, London WC2R 0LR}
\date{}

\title{Explainable AI does not provide the explanations end-users are asking for}

\begin{document}
\maketitle

\begin{abstract}
Explainable Artificial Intelligence (XAI) techniques are frequently required by users in many AI systems with the goal of understanding complex models, their associated predictions, and gaining trust. While suitable for some specific tasks during development, their adoption by organisations to enhance trust in machine learning systems has unintended consequences. In this paper we discuss XAI's limitations in deployment and conclude that transparency alongside with rigorous validation are better suited to gaining trust in AI systems.
\\
\\

\end{abstract}



Artificial Intelligence (AI) is undergoing unprecedented growth across a breadth of applications. These applications range from the trivial, like detecting a face in a picture, to the consequential, like finding a malignant tumor in breast tissue. 

The high importance of some applications is what gives rise to one motivation for the use of explainable artificial intelligence (XAI) systems. Often times end-users of AI systems, rather than AI system developers, seek human-like explanations for their machines' decisions, particularly in high stakes applications like medical diagnoses, legal proceedings, or military operations. The importance of XAI to the end-users is further underpinned by the increasing complexity of modelling techniques. As we will discuss, some predictive algorithms' decisions are easily understood by end-users and practitioners alike, others are not. 

Explanation is defined as ``a statement or account that makes something clear'' or ``a reason or justification given for an action or belief'' in the Oxford dictionary. Not all explanations are equal, however \cite{lipton2001good}. Using ``quantum physics'' as an explanation to the famous double-slit experiment will sufficiently explain the result to one of this article's authors but not the other one as our levels of understanding quantum physics fundamentally differ. Secondly, it is necessary to consider depth of explanation: human circadian rhythm, for instance, can be explained by the day-night cycle we live with. Some, will however not be satisfied with this as an explanation but seek further explanations such as the day-night cycle being result of the rotation of planet Earth. This need for explanation can go further, all the way to the current boundaries of the human knowledge of Physics. Subjectivity is therefore a fundamental limitation of any attempt at explanation \cite{miller2019explanation}. While we struggle to define what exactly the requests for explanations mean we appreciate why these requests are being made.

Central to the end-users desire for explainability is trust. End-users may be more likely to trust machine decisions where said decisions are accompanied by a rationale for why such a decision was arrived upon. Whilst this is a laudable goal, many XAI systems simply do not provide what end-users understand as explanations. Rather, the XAI provides insights into model workings useful for model development. This may lead an XAI system to harm its end-user, rather than help them. Furthermore, explainability as required by end-users is inherently at odds with complex logic required to solve a complex task, simply because it is not possible to communicate explanations of such a high complexity. In this article we argue that XAI is not particularly suitable for establishing trust, and it often provides explanations that will not be considered sufficient by the end users.
Lastly we argue end-users should pay attention to validation, transparency and perhaps interpretability as an alternative to establish trust in AI systems.

\section*{Transparency, Interpretability and Explainability}
\label{sec:Explainability, Interpretability and Transparency}


Transparency, interpretability and explainability are three pertinent terms that are being used almost interchangeably outside of the XAI field, while the differences between them offer a useful angle on the problem.Transparency in particular may offer a more useful alternative for ascertaining trust than explainability. This article adopts the definitions of these terms established by Royal Society \cite{royalsociety2019}.

Transparency refers to how accessible an AI system's internal mechanics, training, and prediction process are to a user \cite{royalsociety2019}. Perfectly transparent AI system will share all source code and data used for training. Without access to training data, independent model validation is unachievable for an end-user \cite{Zhang2022}, resulting in low transparency. Some landmark models in the field \cite{Brown2020, Ramesh2022} do not disclose all parts of the model necessary to recreate them. Lacking transparency can lead to undisclosed model biases \cite{Schramowski2021} and back-door exploitation\cite{Gu2017, Kurita2020}. Altogether the lack of transparency is the cause of reproducibility crisis within the field of AI \cite{Hutson2018}.

A model need not be entirely transparent to be interpretable. Interpretability lies in end-users’ understanding of the underlying principles of the AI system in question \cite{royalsociety2019}. This is unfortunately highly subjective as capability to understand the technology as well as required level of understanding will differ between people.  Neural Networks are an apt example, while the authors of this article may feel they understand the underlying principles, this will not be generally shared across the pool of end-users.


Explainability, meaning how or why a conclusion was reached \cite{royalsociety2019}, is highly subjective too. While other problems in machine learning utilise quantitative methods to make assessments (alike the RMSE of a regression model), no industry standard method to quantify the quality of an XAI explanation has been agreed upon by practitioners. Complex AI systems such as Deep Neural Networks will never be able to provide detailed explanations of how a conclusion was reached simply because of the lack of human capability to comprehend the very large number of parameters and nonlinear interactions. To transform a neural network's prediction into a human explanation therefore requires simplification which in turn will result in sacrificing the accuracy of the explanation. 

 We have introduced the three different concepts characterising different aspects of human understanding of an AI system and argue it is necessary to use these terms precisely as they capture very different qualities of an AI system. Furthermore, we argue that only one of them - transparency - is objectively measurable and therefore suitable for a proper evaluation of an AI system. This article, however, focuses on explainability, which is a very subjective and ill-defined concept fundamentally at odds with complex AI systems. Unfortunately these are not the only issues with ascertaining trust in AI systems through explainability.

\section*{XAI techniques and their uses}
XAI methods can broadly be classified into two categories relating to the form of explanation that they provide: local explainability and global explainability.

Local explainability techniques are used to understand the impact of individual features within a local context of a single prediction. 
Examples of such techniques are LIME\cite{Ribeiro2016LIME}, SHAP\cite{Lundberg2017SHAP} and GradCAM\cite{Selvaraju2016GRADCAM}.

\if{false}
\begin{figure}[h]
    \centering
    \includegraphics[width=100mm]{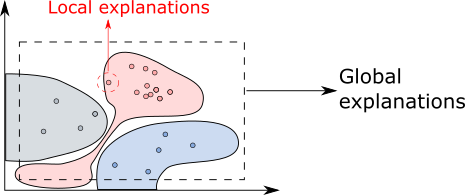}
    \caption{The difference between global and local explanations: the dots represent individual predictions, the shaded area represents the underlying distribution.}
    \label{fig:global_vs_local}
\end{figure}
\fi

Local explanations have also been used to aid the discovery of hidden relationships between features and predictions, one such example was the identification of new radiographic features that predict knee pain which were previously unknown\cite{Pierson2021}. This highlights XAI's most reliable use case: diagnostic tools for AI system developers. These are systems where developers can access quick feedback about feature relevance or even insights into previously misunderstood features resulting in more powerful models \cite{Krishna2022}.

Global explainability techniques aim to explain the average behaviour of the model. This explanation allows some end-users to compare their expected intuition with the model's predictions. However, fidelity is compromised in the absence of a ground truth for comparison. Local interpretations can be aggregated to provide global explanations, however the resulting explanation often depends on the nature of the aggregation \cite{Linden2019}. Some examples of global XAI techniques are Partial dependency plots, Accumulated local effects and Global surrogate models \cite{Linden2019}.

These techniques are often used by AI system developers during the model development phase to understand the overarching relationships that the system is learning. It is useful for AI practitioners to understand the relationships that the model has learnt by explaining its behaviour in terms of extrinsic properties. However, this form of explainability should only be one aspect of gaining trust in these systems, as using explainability tools for the purpose of validating models can actually result in poorly performing models \cite{Krishna2022}.

\section*{Dangers of XAI misuse by end-users}
\label{sec:Dangers of XAI misuse by end-users}

XAI techniques were developed to improve the explainability of machine learning models. However, without a universally accepted way of quantifying the quality of an explanation, data scientists will often use multiple explainability techniques during model development. These explainability techniques can conflict with one another and there is often no systematic way to resolve these differences. This issue is referred to as the disagreement problem of XAI \cite{Krishna2022}. 
Studies performed on commonly used XAI techniques like LIME, KernalSHAP and various gradient methods show that the different explanation techniques disagree with each other on a range of feature importance metrics like rank and even feature direction. A study looking at how various data scientists resolve the disagreement problem \cite{Krishna2022} in practise revealed that most developers employed arbitrary heuristics (such as their favourite method) to choose the final explanation. Often XAI systems will offer multiple explanations and without a systematic way to distinguish between them, the end-user could choose explanations which align with their internal expectations. This contradicts our very notions of the scientific method, akin to researchers only significant results to present in a paper.

Some methods try to evaluate the fidelity/faithfulness of explanations by comparing them to the results of the ground truth explanations \cite{Liu2021SyntheticBenchmarks}. However, for most practical applications, ground truths for explanations are highly unlikely to be available. Furthermore, what is often presented as ground truths are just internal knowledge which can be inconsistent between subject matter experts, poorly grounded in analysis and biased.

Even when explanations are provided by XAI systems, interpreting what they mean can be challenging. One of the challenges is that these explanations are not robust enough to establish trust. A common example, which we will explore, is a medical image classifier which predicts malignant tissue.

\begin{figure}[h]
    \centering
    \includegraphics[width=50mm]{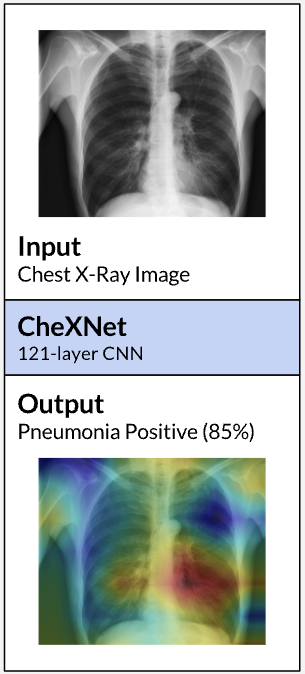}
    \caption{Activation map showing which areas of the lung in a chest X-ray contributed towards the prediction of pneumonia by CheXNet \cite{Rajpurkar2017}}
    \label{fig:chexnet_cam}
\end{figure}

Consider the activation map in Figure \ref{fig:chexnet_cam} showing the explanation for a classifier built to predict if a chest X-ray is showing signs of pneumonia. The activation map highlights which region of the image is important for a prediction. For some practitioners, this explanation will be useless as it doesn't explain how the model came to its classification of pneumonia. It will not aid them in understanding how to improve the performance of the model.

For complex models, the explanations generated by these systems will not be interpretable to end-users. This should be no surprise as humans also cannot explain our own complex decisions. Visual object recognition is a great example, humans are unable to explain recognizing any but the simplest objects in respect to the visual input. Instead we resort to the use of latent variables such as ``the object on the picture has ears'' which themselves are not possible to explain. By the time we reach definitions in terms of actual visual input the explanations become meaningless. Similar issues arise when XAI techniques are used to explain a complex model.
In our experience end-users will struggle to find explanations useful, or worse still, misinterpret them as a result.

To improve trust in some of our solutions, we often share the results of our explainability analysis. On some of our projects, we observed that the simplicity of the output from XAI techniques led to several potential issues typical of XAI systems’ interaction patterns.  Often in these situations, end-users exhibited a preference for using the explainability results over the actual model because of its simplicity and the tendency to reinforce their own beliefs through confirmation bias, despite the XAI results were themselves of significantly lower accuracy than the actual model. Additionally, users on occasion wrongly assumed a causal relationship between the inputs and outputs of the model features.  In this instance, robust validation would be more suitable to support end-user trust.

Anecdotes such as these raise questions about the appropriateness of XAI techniques for end-users.  The quality and appropriateness of these explanations are only some of the challenges associated with using XAI techniques for examining machine learning systems. The goal that these techniques are looking to achieve around trust is conclusively better served through robust validation. Their lack of robustness also leaves them open to many forms of exploitation by malicious actors.

\section*{Vulnerabilities in XAI systems}
\label{sec:Dangers of explainability for end-users}

One way in which an XAI system can be manipulated is known as an adversarial attack \cite{Ghorbani2019}. Adversarial attacks are small and indistinguishable perturbations to the input that usually cause large changes to the output. These methods can also be used to change outputs of XAI without actually removing biases of the underlying model.

Another type of vulnerability that XAI systems are exposed to is fair-washing\cite{Aivodji2019}. A biased black-box model can be rationalised to provide an unbiased explanation creating an illusion of fairness. Given the disagreement between XAI methods a developer can simply select XAI method which did not discover bias and present its results to argue the underlying model is unbiased.

Lastly, like any machine learning model XAI systems can also learn spurious correlations from data without proper validation. Unfortunately most current XAI techniques do not involve any validation at all. Again, a robust and thorough approach to validation is the appropriate solution for gaining trust.

All of these problems highlight the issues with XAI systems which limit their applicability for end-users. These systems are often used to gain trust in a  black-box system, but in the presence of so many vulnerabilities, their performance and output are compromised. We argue for the empiricist's viewpoint whereby all knowledge is obtained through experiences and interactions with the environment. A robust framework for validation and testing is the only way to definitively establish trust in black-box machine learning systems.

\section*{Conclusion}
While explainability can be useful during model development and for monitoring purposes, it is less useful for end-users. A rigorous validation framework is the only acceptable way to assess the performance and biases present in a model. A validation framework, when tailored towards the specific end-user requirements circumnavigates the need for XAI systems.

There are many scientific applications in society where trust is gained through repeated and rigorous validation. Continuing our medical example, consider the process of clinical trials for new drugs. These drugs go through extensive randomised controlled trials, where effects and safety are tested and monitored through progressively staggered trials until deemed safe to use in the general population. Even after release, their efficacy continues to be monitored. By documenting the stages of validation towards the release of this equally black-box system, clinical researchers engender trust in end-users (patients) with no medical subject matter knowledge. Many accepted drugs even have no known explanation for their interactions but the impact is clear, conclusive, beneficial, and most importantly; robustly validated.

We conclude that rigorous validation of AI systems is far more useful to end-users than explainability. Transparency should be pursued to ascertain trust as it enables validation to be performed independently of the creators of AI system in question.

\section*{Acknowledgments}
This work was supported by Arca Blanca. We are immensely grateful for the help Nelson Peace provided to make this article user-friendly and ready for publication.

\bibliographystyle{unsrt}  
\bibliography{references}

\end{document}